\documentclass[]{spie}  
\usepackage[]{graphicx}

\title {Study of the scientific potential of a three 40 cm Telescopes
  Interferometer at Dome C.} 

\author{B. Valat\supit{a},  F.X Schmider\supit{a},
  B. Lopez\supit{b},R. Petrov\supit{a}, M. Vannier\supit{c},
  F. Millour\supit{d}\supit{a}, F. Vakili\supit{a}.
\skiplinehalf
\supit{a}Laboratoire Universitaire d'Astrophysique de Nice, Parc Valrose, Nice, France; \\
\supit{b}Observatoire de la C\^ote d'Azur, Gemini, Boulevard de l'Observatoire,
Nice, France;\\
\supit{c}European Southern Observatory, Chile;\\
\supit{d}Laboratoire d'Astrophysique de l'Observatoire de Grenoble,
  Grenoble, France;
}

\authorinfo{Further author information: E-mail: bruno.valat@unice.fr, Telephone: 33 (0)4 92 07 61 98\\}

\begin{document} 
  \maketitle 

\begin{abstract}

Recent site testing (see:
http://www-luan.unice.fr/Concordiastro/indexantartic.html) has shown
that Dome C in Antarctica might have a high potential for stellar
interferometry if some solutions related to the surface atmospheric
layer are found. A demonstrator interferometer could be envisioned
in order to fully qualify the site and prepare the future development
of a large array.
	
We analyse the performances of a prototype interferometer for Dome C
made with 3 telescopes of 40 cm diameter. It assumes classical
Michelson recombination. The most recent
atmospheric and environmental conditions measured at Dome C are
considered (see K. Agabi  "First whole atmosphere night-time seeing
measurements at Dome C, Antarctica"[\cite{Agabi}]). We also study the possible
science reachable with such a demonstrator. Especially we evaluate
that even such small aperture interferometer could allow the detection
and low resolution spectroscopy of the most favourable pegaside
planets.

\end{abstract}


\keywords{long baseline interferometry, astronomical sites: Antarctica Dome C, extrasolar planets}

\section{INTRODUCTION}
\label{sect:intro}  
Astronomical observations in infrared are mainly limited by the atmospheric
background. This flux depends strongly on the atmospheric
temperature and emissivity. That is why the Antarctic should be an interesting place
for astronomical observations (Chamberlain, 2000 [\cite{astoo}]). Since the
forthcoming of permanent stations in Antarctica, more and more
research groups are testing atmospheric characteristics there and expect
to build science instruments on Antarctic plateau. 
The size of the projects ranges begin from a small telescope to an ELT
or to an optical interferometric array like Kiloparsec Explorer for
Optical Planet Search (KEOPS)(Vakili, 2005 [\cite{KEOPS}]). Since a few years the
Laboratoire Universitaire Astrophysique de Nice (LUAN) has
access to the station Concordia. The site
characteristics were investigated. A strong
potential of this site for interferometry was pointed out.

\section{Dome C, one of the most promising earth site}
\subsection{Site description}
The Concordia station is located at Dome C, at an altitude of 3300
meters. Since few years the LUAN has been characterizing the atmosphere at
Dome C in order to know the real potential of this location. The first
results of these tests present Concordia as an exceptional site for
astrophysics both for its reduced atmospherics background and for its
low atmospheric turbulence during the summer days.

The 210$^o$ Kelvin temperature during the winter night implies for this site a very
low atmospheric thermal background.
The weather is very favorable to astronomical observations with a
clear sky 96\% of the time (see E. Aristidi, 2005
[\cite{Aristidi}]). These properties show that the Dome C has the
clearest atmosphere on earth. This imply the possibility to
observe faint objects in infrared more easily than in the other sites.
Moreover the background fluctuations should have a low amplitude. As the
background fluctuation is correlated to the water vapor lines
(Miyata, 1999 [\cite{Miyata}]), the low water concentration at Dome C should imply a
low background fluctuation.

\begin{figure}
   \begin{center}
   \begin{tabular}{c}
   \includegraphics[height=7cm]{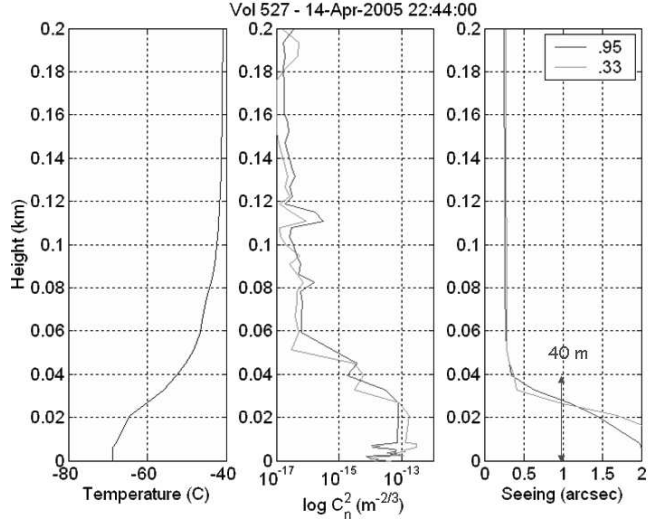}
   \end{tabular}
   \end{center}
   \caption[boundelayer] 
   { \label{fig:boundelayer}: C $^2_{n}$ and temperature and the
   derived seeing as function as the altitude [\cite{Aristidi}].}
   \end{figure}

The sky brightness is not the only advantage of this site.
The seeing of this site is very good too. During summer the mean
seeing is 0.2 arcsec and during the winter the value is 1.4 arcsec at
the ground level [\cite{Agabi}]. As
shown in the figure \ref{fig:boundelayer}, the atmospheric turbulence is
principally located in the first 30 meter. The 30m high seeing calculated
thanks to balloons measurements is of the order of 0.4 arcsec during
the winter. In order to reach
this low seeing two ways are studied: a ground adaptive optic system
or elevating the telescopes over the turbulence layer. 

All these results present the Dome C as an exceptional site for
``classical'' telescopes but it has also interesting characteristics for
interferometers. The coherence time is 6.8 ms; It is
twice as large as the one on the best actual interferometer site
(Paranal). Moreover the atmosphere at Dome C has a low outer scale (and a high inner
scale) which implies that the piston should be small. 
The last important parameter of an interferometer site is the
isopistonic angle $\theta_{0}$. $\theta_{0}$ at Dome C is 6.8 arc
sec, which is 3 times larger than the one at Paranal or 2 times upper
than the one at Maidanak. Such isopistonic angle will
help to find a reference object needed for faint
observations. In the following proceeding, the only thermal noise and the
readout noise were considered in our evaluations.

\section{background simulations and expected science with a three 40 cm
interferometer}

\subsection{Thermal background simulated for Dome C}
In order to evaluate the real potential of an instrument at Dome C, we
simulated the fundamental noises. First, the sky and instrument thermal
backgrounds are presented. The flux from the thermal emission is
simulated thanks to a black body emission. The simulated telescopes
are assumed to be perfect, vibrations and the residual error from fringe
tracking are neglected. The field of view of the system is given
by the following equation:

	\begin{equation}
	\label{eq: numerical aperture of the tlescop }
NA  =   (1.22 \lambda / D)^2 \pi^2 (D)^2 ,
	\end{equation}
Where NA is the field of view $\lambda$ the working wavelength and
D is the telescope diameter.

\begin{figure}
   \begin{center}
   \begin{tabular}{c}
   \includegraphics[height=7cm]{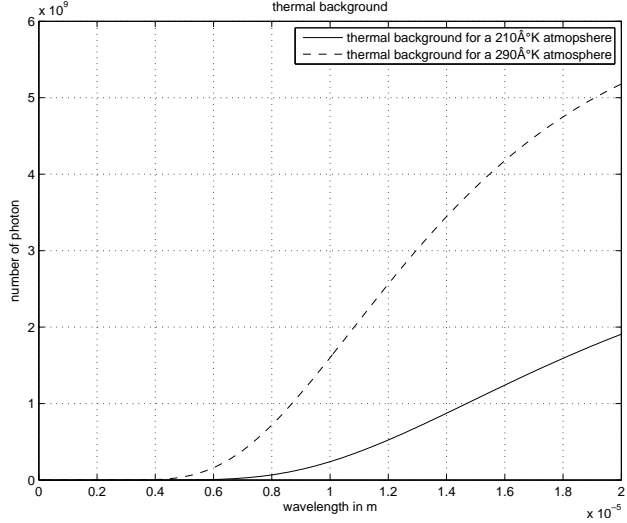}
   \end{tabular}
   \end{center}
   \caption[background] 
   { \label{fig:background}: Thermal background calculated for Dome C
   (210$^o$K) and Paranal (290$^o$K) conditions.}
   \end{figure}

The thermal emission of the atmosphere and the telescope is compared
on figure \ref{fig:background} with the background of a same system at 290$^o$K
(typically the temperature of the atmosphere at Paranal). These
simulations show that the atmospheric background level at Dome C is 3
times fainter than the one measured at Paranal.

In order to be able to compare the background flux to a science
source and hot Jupiter signal, a flux calculation was carried out and
exposed in the next subsection.

\subsection{Flux from a star and a hot Jupiter received by a 40cm
  telescopes interferometers, differential phase closure measurement}

A black body emitting at the stellar temperature simulates the star
emission. The planet emission is simulated by the sum of a black body emitting at
the planet temperature and the reflection of the star flux on the
planet surface. 

The modelling of the planet emission with a black body is a rough
approximation. This approximation is pessimistic in Near infrared
(J,H,K) but in far infrared the simulations should be carried out with
other atmospheric model [Barman, 2000 [\cite{barman}]).
The albedo of the hot Jupiter is around 0.1. The position of the planet was
taken as the most favorable, and we consider that one half of the planet receives
and reflects the light from the star.
For this article three hot Jupiter's were selected and simulated. Their
characteristics are exposed in the following tables.

\begin{table}[h]
\caption{Property of the host stars for the hot Jupiter systems of interest} 
\label{tab:fonts}
\begin{center}       
\begin{tabular}{|c|c|c|c|c|c|c|} 
\hline
\rule[-1ex]{0pt}{3.5ex} star & system & spectral & K mag & V mag & star	& star\\
 & distance (parsec) & type & &	& diameter (km) & temperature ($^o$K) \\
\hline
\rule[-1ex]{0pt}{3.5ex}  HD108147 & 38.5 & F8 & 5.72 & 6.98 & 970850.1 & 6130 \\
\hline
\rule[-1ex]{0pt}{3.5ex}  HD13445 (GJ86) & 10.9 & K1 & 4.13 & 6.12 & 650274.9 & 5350 \\
\hline
\rule[-1ex]{0pt}{3.5ex}  HD179949 & 27 & F8 & 4.94 & 6.24 & 983249.7 & 6100 \\
\hline
\end{tabular}
\end{center}
\end{table} 

The stars radii are calculated thanks to the following empiric formula
(Kervella, 2004 [\cite{Kervella}]):
	\begin{equation}
	\label{eq: Angular diameter relations based on color }
\log \theta_{LD} = c_{\lambda} (C_{0}-C_{1})+ d_{\lambda} - 0.2 C_{0} ,
	\end{equation}

Where $\theta_{LD}$ is the angular diameter of the star in mas, $C_{0}$ and $C_{1}$ are any
two distinct colors of the Johnson system, and $c_{\lambda}$ and $d_{\lambda}$
are two factor depending on the spectral band of $C_{0}$ and $C_{1}$.

\begin{table}[h]
\caption{Characteristics of the interesting hot Jupiter} 
\label{tab:fonts}
\begin{center}       
\begin{tabular}{|c|c|c|c|c|} 
\hline
\rule[-1ex]{0pt}{3.5ex} star	& separation 	& planet 		& planet	& planet\\
				& (AU) 		& temperature ($^o$K)	& mass(M$_{Jup}$)& diameter (km)\\
\hline
\rule[-1ex]{0pt}{3.5ex}  HD108147 & 0.104 & 890 & 0.4 & 200177\\
\hline
\rule[-1ex]{0pt}{3.5ex}  HD13445 (GJ86) & 0.11 & 800 & 4.01 & 157282 \\
\hline
\rule[-1ex]{0pt}{3.5ex}  HD179949 & 0.04 & 1160 & 0.98 & 150133,2 \\
\hline
\end{tabular}
\end{center}
\end{table} 

\begin{figure}
   \begin{center}
   \begin{tabular}{c}
   \includegraphics[height=7cm]{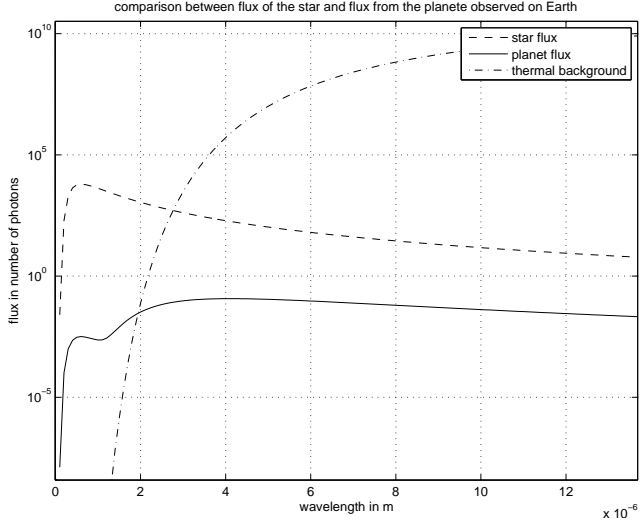}
   \end{tabular}
   \end{center}
   \caption[stellar flux] 
   { \label{fig:fluxsample}: The planet, star and thermal background
   flux received by a 40cm telescope based on earth. The simulated system is
   HD108147 with a exposure time of 10ms, the transmission is 10\%}
   \end{figure}

The planet diameter is calculated thanks to the empiric formula given by
Guillot (Guillot, 1999 [\cite{Guillot}]).
The
flux resulting from the simulation for the less favorable case
(HD108147) are plotted on the figure \ref{fig:fluxsample}.

The planet detection can be carried out thanks to the differential
method (Vannier, 2006 [\cite{Vannier}]). In order to know the best spectral band to
detect hot Jupiter the signal to noise ratio should be calculated.
The fundamental limitation in the hot Jupiter detection is the thermal
background emission and the readout noise.

\subsection{Signal to noise ratio for hot Jupiter detection}

In the previous subsection we saw that the highest part of the
planetary signal is overflowed by the thermal background. 
The thermal noise should be calculated in order to define the best
working wavelength and the observation time needed to reach a good
signal to noise ratio (Vannier, 2006 [\cite{Vannier}]).
The phase error from fundamental noise is (\ref{eq:phasenoise})[\cite{Vannier}].

 	\begin{equation}
	\label{eq:phasenoise}
\sigma =\sqrt{3} \frac{\sqrt{((\sigma^2_{phot} + M np \sigma^2_{read} + \sigma^2_{th})/2)}} {(C
N(\lambda_{i})/n_{tel})}
	\end{equation}

Where N($\lambda_{i}$) is the total number of photon per channel $\lambda_{i}$,
$\sigma^2_{phot}$ is the photon noise, $\sigma^2_{read}$  is the read out noise, and $\sigma^2_{th}$ is the variance of the thermal noise.
M is the number of frames, np the number of pixel per spectral
channel, C the contrast and n$_{tel}$ the number of telescopes.

The signal to noise ratio defines that the best window  for a three
40cm telescope interferometer is the K band. The simulated
interferometer has a 200 m baseline, the exposure time is 10ms, and
the $\delta\lambda$ of 0.1 $\mu$meter.
For comparison the integration time needed to reach a signal to noise ratio over 3
in the most favorable case: GJ86 is 10 h and for the farest: HD179949 is 200 h
and for HD108147 is 600 h. The integration time needed to reach such
signal to noise ratio with AMBER is for GJ86 1h, for HD179949 3h, and
for HD108147 50h. 

If just the thermal background gain is took into account, the signal to
noise ratio of an interferometer composed by three 8 meter telescope
at Paranal will be equivalent to one composed by three 5 meter
telescopes at Dome C. 
As the seeing is better at Dome C, the adaptive optics of a
telescope at Dome C should be better than the one at Paranal, and the
total flux transmissions should be better. The transmission could be 3
times better for an interferometer at Dome C than the same at Paranal,
the 8 m Paranal telescope are corresponding to a 3.5 m Dome C
telescope in the case of differential color measurement.

\begin{figure}
   \begin{center}
   \begin{tabular}{c}
   \includegraphics[height=7cm]{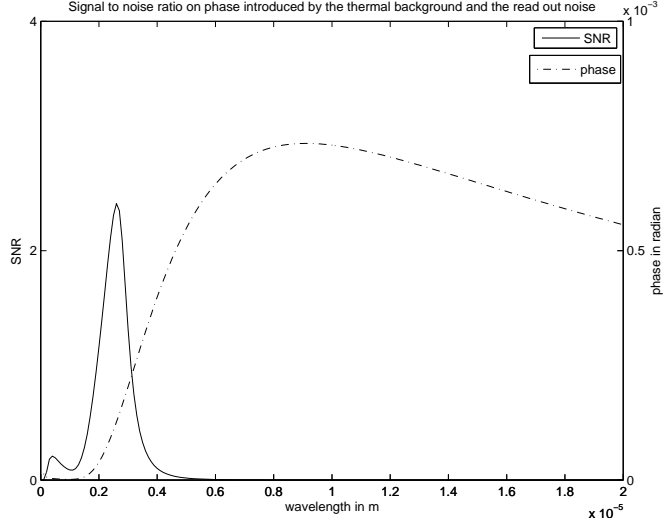}
   \end{tabular}
   \end{center}
   \caption[SNR] 
   { \label{fig:SNRsample}: Signal to noise ratio for an closure phase
   on HD108147 with an exposure of 600 hours.}
   \end{figure}

The signal to noise ratio of the HD108147 is plotted on the figure \ref{fig:SNRsample}. 

As the Dome C atmosphere is stable enough to allow a 10h integration time
observation of some hot Jupiter should be possible.

\section{Myk\'erinos an interferometer prototype at Dome C}

Our laboratory propose to develope an optical interferometric array:
KEOPS.(Vakili 2005 [\cite{KEOPS}]) This interferometer will be composed by thirty six 1.5 meters
telescopes deployed on 3 concentric rings with a maximal diameter of
1 km. Before building such complex interferometer the creation of a
smaller interferometer would be carried out in order to validate
the technical choices: Myk\'erinos.
Myk\'erinos will be composed by three 40 cm telescopes with a 200 meters
baseline. It will work in phase closure in order to remove
instrumental and atmospheric phases (Segransan 1999 [\cite{segransan}]
Danshi, 2006 [\cite{Danshi}]).
As shown on the last subsection this prototype is very promising for
astrophysics. Such small interferometer can already study hot
Jupiter, high contrasted binaries, and measure fundamental parameters like star
diameters, or star masses.

\subsection{Myk\'erinos Challenges}

This interferometer will allow the confirmation of technical design for future
Antarctica interferometer. Working in critical locations like Dome C,
requires a high system automation for avoiding human
interactions. The present logistics does not allow the presence of
large team at Dome C and the temperature over winter do not allow
instrumental adjustments. The system should be able to work at low
temperature (190$^o$K) with a low water concentration.
An other critical technical point for this interferometer is to define the best
way to get rid of the ground layer. Two solutions are actually considered : using
ground layer adaptive optics or raising the telescopes on the top of 30
meter towers.

\subsection{Myk\'erinos design and forthcoming step}

Myk\'erinos will be composed by three 40 cm telescopes.
The maximal base line is 200 meters.
The flux from the telescopes are injected in fibers in order to be transported to
the delay lines. The main problem is the chromatic dispersions
which should be studied in order to find a compromise (Vergnole, 2005 [\cite{Ohana}]).
According to the turbulence measured by Aristidi (Arisitidi, 2005 [\cite{Aristidi}]) if the telescopes are over 30m
towers, a tip tilt mirror should be enough to remove the effect of the
atmosphere from visible to 2 $\mu$m.
The beam combiner could be integrated optics techniques in order to
have a high efficiency (Lebouquin, 2006 [\cite{integrated}]).

\bibliography{mykerinosarticle}
\bibliographystyle{spiebib}

\end{document}